\documentclass[%
 reprint,
 amsmath,amssymb
 aps,
]{revtex4-1}

\usepackage{graphicx}
\usepackage{dcolumn}
\usepackage{bm}
\usepackage{color}
\usepackage{wasysym}
\usepackage{epstopdf}
\usepackage{verbatim}
\usepackage{epsfig}

\begin{document}

\preprint{APS/123-QED}

\title{Laser cooling of $^{85}$Rb atoms to the recoil-temperature limit}

\author{Chang Huang, Pei-Chen Kuan}
\author{Shau-Yu Lan}%
 \email{sylan@ntu.edu.sg}
\affiliation{%
Division of Physics and Applied Physics, School of Physical and Mathematical Sciences, Nanyang Technological University, Singapore 637371, Singapore
}%




\date{\today}

\begin{abstract}
We demonstrate the laser cooling of $^{85}$Rb atoms in a two-dimensional optical lattice. We follow the two-step degenerate Raman sideband cooling scheme [Kerman \emph{et al}., Phys. Rev. Lett. {\bf84}, 439 (2000)], where a fast cooling of atoms to an auxiliary state is followed by a slow cooling to a dark state. This method has the advantage of independent control of the heating rate and cooling rate from the optical pumping beam. We operate the lattice at a Lamb-Dicke parameter $\eta$=0.45 and show the cooling of spin-polarized $^{85}$Rb atoms to the recoil temperature in both dimension within 2.4 ms with the aid of adiabatic cooling.

\end{abstract}

\pacs{Valid PACS appear here}
\maketitle
Laser cooling technologies \cite{chu,metcalf} of neutral alkali-metal atoms have enabled the high-precision test of the fundamental physics \cite{Cronin,Ludlow}, a study of atom-light interaction at the level of single quanta \cite{Hammerer}, advances in quantum metrology \cite{Ludlow}, etc. Various steps of laser cooling methods, such as Doppler cooling, polarization-gradient cooling, or optical lattice cooling are employed for different purposes in experiments. In particular, the Raman sideband cooling of neutral atoms to the ground state of a harmonic potential in an optical lattice has been a dependable method used to achieve both high phase space density and large atom numbers compared to other cooling methods. It has been used to pre-cool atoms rapidly for the production of a Bose-Einstein condensate by a subsequent evaporative cooling. Recently, the formation of a Bose-Einstein condensate only by Raman sideband cooling has been demonstrated by further improving the density of atoms during the cooling cycle \cite{hu}. In experiments of precision measurement and quantum optics, the preparation of ultralow temperature and a large number of samples is indispensable to obtain a large signal-to-noise ratio and a long coherence time.

Sideband cooling was first demonstrated with ions \cite{Diedrich}. In neutral atoms, Raman sideband cooling has been demonstrated with $^{6}$Li \cite{Parsons}, $^{39}$K \cite{Gröbner}, $^{87}$Rb \cite{hu,kerman2,thompson,kaufman}, and $^{133}$Cs \cite{kerman2,perrin,hamann,vuletic,kerman,han}. Different experimental arrangements are required for different fine and hyperfine structures of atomic species. $^{87}$Rb atoms are of great interest in atomic physics due to their high natural abundance in Rb isotopes and opposite sign of scattering length of their hyperfine ground states \cite{Kokkelmans}. They have been used to study the scattering properties of cold atomic mixtures \cite{li2,altin,papp,deh,cho} and quantum optics experiments \cite{lan,lan2}. There are also ongoing experiments to test Einstein's equivalence principle with $^{85}$Rb and $^{87}$Rb atom interferometers either on the ground or in space \cite{zhou}. On the other hand, the small energy separation of the hyperfine states and large collision loss make $^{85}$Rb less popular than $^{87}$Rb in the community. Here, we demonstrate the degenerate Raman sideband cooling of $^{85}$Rb atoms in a two-dimensional optical lattice with about 90$\%$ of the initial atoms loaded into the lattice.

\begin{figure}
\includegraphics[scale=0.3]{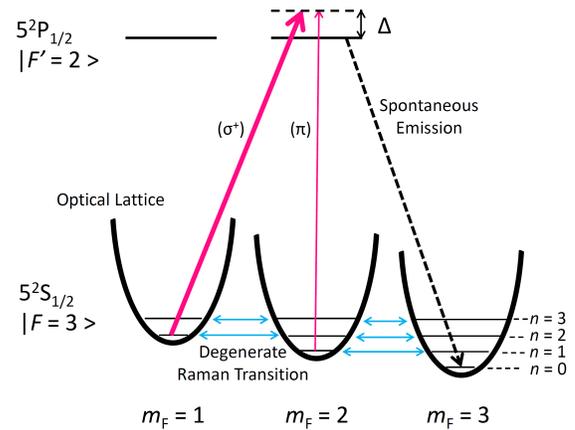}
\caption{\label{fig:epsart} Two-step degenerate Raman sideband cooling energy scheme. The figure only shows the most relevant energy levels. The blue double arrows indicate the degenerate two-photon Raman transition for the interchange of vibrational energy and Zeeman energy. The dashed arrow indicates the relevant spontaneous emission in the cooling process. Due to the small separation between the excited hyperfine states of the $D_{2}$ line manifold, we employ a $D_{1}$ line for optical pumping to avoid coupling to other excited states. The $\sigma^{+}$-polarized optical pumping beam (thick red arrow) represents the fast cooling to $|F$=3, $m_\textrm{F}$=2, $n$=0$\rangle$ state, and the $\pi$-polarized optical pumping beam (thin red arrow) represents the slow cooling to the $|F$=3, $m_\textrm{F}$=3, $n$=0$\rangle$ state. Both frequencies are detuned by $\Delta$ from the excited state. The light shift and power broadening from the optical pumping beam and the lattice beams are omitted in the figure.}
\end{figure}

One of the particular Raman sideband cooling schemes in an optical lattice is to implement an external magnetic field to shift the adjacent Zeeman states by one vibrational quanta of the lattice harmonic potential. The vibrational energy of the atoms can then be transferred to their Zeeman internal energy via an off-resonant degenerate two-photon Raman transition induced by optical lattice beams. The Zeeman internal energy is then removed through an optical pumping process where atoms at the end are shelved into a dark state to decouple them from the optical pumping and two-photon Raman process. To ensure atoms remain in the same vibrational level after optical pumping, the whole process has to operate in the Lamb-Dicke regime, i.e., $\eta=\sqrt{E_{R}/\hbar\omega}<1$, where $\hbar$ is the reduced Planck constant, $E_\textrm{R}$ is the recoil energy, and $\omega$ is the angular vibrational frequency of the harmonic potential. These criteria and the "darkness" of the final dark state set the limit on the lowest temperature that can be obtained. Figure 1 shows the cooling steps and the relevant energy levels of $^{85}$Rb in one of the dimensions for the experiment. The degenerate Raman coupling between Zeeman states $m_\textrm{F}$ and $m_\textrm{F-1}$ is induced by the lattice beams. In the pioneering two-step cooling scheme \cite{kerman2,kerman}, atoms in high lying vibrational states are rapidly accumulated in the auxiliary state $|F$=3, $m_\textrm{F}$=2, $n$=0$\rangle$ by a strong $\sigma^{+}$-polarized optical pumping beam and Raman coupling. Due to the linear Zeeman shift, atoms in this state are decoupled from other lower Zeeman states in the two-photon Raman process. A weak $\pi$-polarized optical pumping beam then puts atoms favorably into $|F$=3, $m_\textrm{F}$=3, $n$=0$\rangle$ by spontaneous emission. The $\pi$-polarized optical pumping beam is weak enough to ensure the vibrational levels are well resolved and minimize heating from single photon scattering and light-assisted collision. The frequencies of the optical pumping beam are chosen on the $F$=3 to $F'$=2 transition frequency such that $|F$=3, $m_\textrm{F}$=3, $n$=0$\rangle$ can only couple to excited states off resonantly.

\begin{figure}
\includegraphics[scale=0.28]{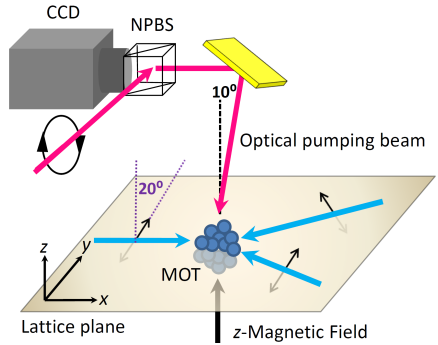}
\caption{\label{fig:epsart} Optical arrangement for the experiment. The thick blue arrows on the horizontal $x$-$y$ plane are the lattice beams and the red arrow from the top is the optical pumping beam. The imaging beams are from the cooling and repumping beams of the MOT and the florescence images are collected from the CCD camera. NPBS is a non-polarizing beamsplitter for separating the optical pumping beam and fluorescence.}
\end{figure}

\begin{figure}
\includegraphics[scale=0.33]{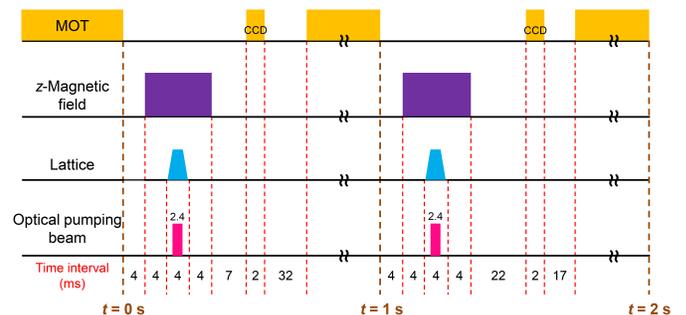}
\caption{\label{fig:epsart} Timing sequence. The sub-Doppler cooling process ends at $t$=0 s and the $t$=1 s, where MOT cooling, repumping beams, and anti-Helmholtz are off. Each experimental trial lasts for 2 s. Each second differs in the imaging time for the time-of-flight measurement. The optical pumping beam is on for 2.4 ms. The optical lattice power is ramped up and down in 800 $\mu$s. The sideband cooling duration in the diagram includes only the time where the optical pumping beam is on.}
\end{figure}

Our two dimensional optical lattice is formed by three coplanar optical beams intersecting at an angle of 120$^{\circ}$ as shown in Fig. 2. The lattice beams are generated from a tampered amplifier seeded by an extended cavity diode laser. The diode laser is injection locked by the magneto-optical trap (MOT) cooling beam which is modulated by an electro-optic modulator at 12 GHz. The negative first-order sideband of the MOT beam after the electro-optic modulator is filtered by a temperature-stabilized solid \'{e}talon and injected into the diode laser. An external magnetic field is pointing along the normal vector of the lattice plane to define the quantization axis for atom-light interactions. The optical lattice beams are linearly polarized and the polarization is tilted by about 20$^{\circ}$ with respect to the quantization axis to provide a nonzero transition amplitude of the two-photon Raman transition. When the tilting angle is small, we can approximate the lattice potential as

\begin{equation}
\begin{aligned}
U=&-\frac{4}{3}u\bigg[\frac{3}{2}+\cos(\sqrt{3}ky)+\cos\left(\frac{\sqrt{3}ky-3kx}{2}\right) \\
&+\cos\left(\frac{\sqrt{3}ky+3kx}{2}\right)\bigg],
\end{aligned}
\end{equation}
where \textit{u} is the single beam potential, \emph{k} is the wavevector of the light, and $x$ and $y$ are spatial coordinates \cite{Tai}. By considering a large trap depth compared to the recoil energy, the potential can be approximated further as a harmonic potential around $x$=0 and $y$=0 as

 \begin{equation}
U\approx-u[-6+3k^{2}(x^{2}+y^{2})],
\end{equation}
with oscillation frequency $\omega=\sqrt{\frac{6uk^{2}}{m}}$, where $m$ is the mass of the particle. The temperature $T_{0}$ of the ground state of the harmonic potential can be inferred from $k_{\textrm{B}}T_{0}=\hbar\omega/2$, where $k_{\textrm{B}}$ is the Boltzmann constant.
Our lattice beams have a waist of 7 mm and 47 mW on each beam, and the frequency is red-detuned 12 GHz from the $^{85}$Rb $D_{2}$ line $F$=3 to $F'$=4. The calculated vibrational frequency $\omega$/2$\pi=$21 kHz which corresponds to a temperature $T_{0}$=440 nK.

Atoms are first loaded from the background Rb vapor into a three-dimensional MOT for about 900 ms in a commercial ultrahigh vacuum chamber (miniMOT from ColdQuanta, Inc.). The MOT coils are then switched off to perform sub-Doppler cooling for 30 ms. Through a time-of-flight measurement, the temperature of the atomic ensemble at this stage is about 10 $\mu$K. We conduct a microwave spectroscopy on the Zeeman sublevels and minimize the energy splitting between Zeeman states by applying external magnetic fields from three pairs of Helmholtz coils. The ambient magnetic field is reduced to about less than 10 kHz on the adjacent Zeeman sublevels. To avoid heating, we adiabatically ramp up the power of the lattice beams in 800 $\mu$s to load atoms into the optical lattice. Meanwhile, a tens of mG of magnetic field for defining the quantization axis and to split the Zeeman degeneracy is turned on before switching on the lattice beams.

The optical pumping beam with a beam waist of 7 mm and total power of 1 mW is generated from another extended cavity diode laser at 795 nm and propagating along the direction 10$^{\circ}$ from the quantization axis. The polarization of the optical pumping beam is circular and the offset angle from the $z$ axis gives a ratio of 5 to 10$\%$ on the $\pi$-polarized component and $\sigma^{+}$-polarized component. The frequency of the optical pumping beam is detuned by $\Delta$ from the $D_{1}$ line $F$=3 to $F'$=2 transition. Even with a small probability, atoms are pumped to the $F$=2 state and leave the cooling process. We use the MOT repumping beams to pump atoms back into the $F$=3 state to continue the cooling process. We then turn on the MOT cooling and repumping beams at 7 and 22 ms after the sideband cooling and collect the fluorescence by a CCD camera from the top of the lattice plane. We fit the images with a single Gaussian function and the fitted 1/$e^{2}$ widths are used to calculate the temperature of the atomic cloud through the equation of ballistic expansion. The error bar of each data point of temperature measurement throughout this article only takes into account the fitting error and is about the size of the symbol. Any deviations from the theoretical fittings are due to the time-dependent drift of the optical dipole potential or the ambient magnetic field. The timing sequence is shown in Fig. 3.

\begin{figure}
\includegraphics[scale=0.3]{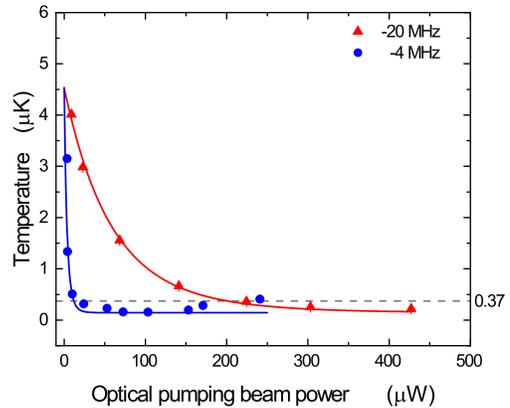}
\caption{\label{fig:epsart} The final temperature in the $y$ axis as a function of optical pumping beam total power with different detuning $\Delta$. The blue circles and red triangles correspond to -4 MHz and -20 MHz detuning from the $D_{1}$ line $F$=3 to $F'$=2 transition, respectively. The lattice beams have a waist of 7 mm and 47 mW on each beam and the cooling process is held for 2.4 ms. The dashed line indicates the recoil temperature of $^{85}$Rb. The curves are the theoretical fitting discussed in the text.}
\end{figure}

In Fig. 4, we measure the temperature as a function of optical pumping beam power at different optical pumping beam detunings. The data are taken with 21 kHz of vibrational energy and 1 mW of optical pumping beam power. We fit the data with an energy rate equation which is the sum of the sideband cooling rate and heating rate from the recoil during the optical pumping process \cite{kerman2,Belmechri}. The curve fitting also takes into account the adiabatic cooling by ramping down the lattice power in 800 $\mu$s \cite{Kastberg}. Although the $\sigma^{+}$-polarized optical pumping beam at -4 MHz detuning causes larger energy shift (tens of kHz) and broadening (tens of kHz) on the energy levels than -20 MHz detuning, the final temperature does not depend on the detuning. This is due to the decoupling of the $|F$=3, $m_\textrm{F}$=2, $n$=0$\rangle$ and $|F$=3, $m_\textrm{F}$=3, $n$=0 and 1$\rangle$ states from the strong $\sigma^{+}$-polarized optical pumping beam and is only addressed by the weak $\pi$-polarized optical pumping beam. The increase of the temperature over a large optical pumping beam power at -4 MHz detuning is due to radiation pressure pushing atoms away from the lattice area. We investigate this increase of temperature in Fig. 5.

\begin{figure}
\includegraphics[scale=0.3]{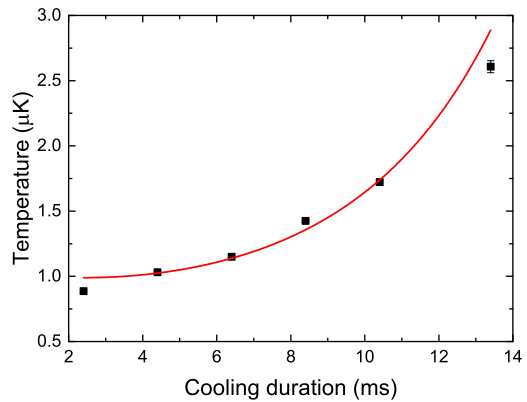}
\caption{\label{fig:epsart} The final temperature after sideband cooling as a function of  the cooling duration. The total optical pumping beam power is 1 mW and the detuning $\Delta$=+20 MHz from $|5S_{1/2},F=3\rangle$ to $|5P_{1/2}, F'=2\rangle$ transition. Each lattice beam power is 47 mW. The measurements are done in the $y$ axis.}
\end{figure}

Figure 5 shows the temperature of the atomic ensemble with different cooling durations. Each experimental data point is an average of 30 experimental trials. Due to the arrangement of the optical pumping beam (see Fig. 2), atoms receive a momentum kick of $\hbar$$\textbf{k}_{op}$ in the direction orthogonal to the lattice plane for each scattering event and, therefore, will drift out of the high trapping potential area for a longer cooling duration, where $\textbf{k}_{op}$ is the wavevector of the optical pumping beam. We fit the data with an energy rate equation as in the fitting of Fig. 4. We assume the sideband cooling process reaches equilibrium at 2.4 ms in which atoms stop receiving momentum kicks in the optical pumping beam direction. Atoms start to drift out of the lattice area at a constant initial velocity at 2.4 ms and the vibrational energy is set as a function of the time in the fitting. The increase of the temperature is mainly due to the adiabatic cooling process which works more efficiently at a larger vibrational energy \cite{Kastberg}.

\begin{figure}
\includegraphics[scale=0.3]{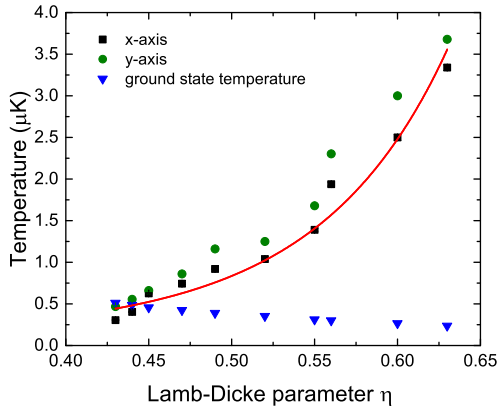}
\caption{\label{fig:epsart} The final temperature with varying optical potential in the $x$ and $y$ directions. The horizontal axis is the corresponding Lamb-Dicke parameter. The blue triangles are the calculated temperature of the ground-state energy of the harmonic potential for reference. The fitting curve is for the $x$ axis only.}
\end{figure}

In Fig. 6, we measure the temperature of atoms with different Lamb-Dicke parameters (different lattice beam power) in two orthogonal directions, $x$ and $y$. At each Lamb-Dicke parameter, we adjust the $z$-axis magnetic field to optimize the temperature. The cooling duration are kept at 2.4 ms and the optical pumping beam power is 1 mW at +20 MHz detuning. We compare the measured temperature with the calculated ground state  temperature at different Lamb-Dicke parameters. The atoms' temperature starts to approach the ground state temperature at around $\eta$=0.45 in our experiment. The red curve shows the theoretical fitting of the $x$-axis measurement. The fitting uses the same energy rate equation as in Fig. 4 but sets $\eta$ as the fitting variable. Due to the imperfection of the alignment of the lattice beams, the two axes have a slightly different lattice potential which leads to different final temperatures. Besides the Lamb-Dicke parameter, the cooling efficiency is limited by how dark the final state can be. The single-photon scattering from the lattice beams and the $\pi$-polarized optical pumping beam could pump atoms out of the $|F$=3, $m_\textrm{F}$=3, $n$=0 $\rangle$ state. A light shift and broadening of the optical pumping beam could cause an off-resonant Raman transition which heats up instead of cools down the atoms or couples atoms out of the dark state \cite{kerman2}.

We estimate the density of the atomic ensemble by measuring the optical depth (OD) after sideband cooling. A circularly polarized probe beam of 1 $\mu$W and 2 $\mu$s long resonant on the $D_{2}$ line $F$=3 to $F'$=4 is sent along the quantization axis direction. We measure the transmission of the probe beam and obtain OD=7(2). Combining the size of the ensemble $L$=0.7(2) mm in the $z$ direction, we calculate the number density of atoms $n=8.0(3)\times 10^{10}$cm$^{-3}$ after sideband cooling from OD$=\sigma nL$, where $\sigma$ is the resonant cross section for isotropic light polarization. The corresponding phase space density $n\lambda_{\textrm{dB}}^{3}$ is about 1/581, where $\lambda_{\textrm{dB}}$ is the thermal de Broglie wavelength.

In summary, we demonstrate the laser cooling of optically trapped $^{85}$Rb atoms in a harmonic potential within 2.4 ms. The final temperature is consistently around the recoil temperature $T_{r}$=370 nK and the number of atoms after cooling is 1.4(2)$\times$ 10$^{7}$. It can be extended to three-dimensional cooling by adding another lattice beam in the $z$ axis. This low temperature atomic ensemble could be used to enhance other cooling methods such as evaporation cooling and delta-kick cooling in time and total atomic numbers for quantum optics and fundamental physics experiments with $^{85}$Rb.

This work is supported by Singapore National Research Foundation under
Grant No. NRFF2013-12, Nanyang Technological University under start-up grants, and
Singapore Ministry of Education under Grants No. Tier 1 RG193/14.


\nocite{*}

\bibliography{apssamp}

\begin{thebibliography}{00}
\bibitem{chu}Steven Chu, Rev. Mod. Phys. {\bf70}, 685 (1998); ibid. {\bf70}, 707 (1998); ibid. {\bf70}, 721 (1998).
\bibitem{metcalf}H. Metcalf and P. van der Straten, Laser Cooling and Trapping (Springer, New York, 1999).
\bibitem{Ludlow}Andrew D. Ludlow, Martin M. Boyd, Jun Ye, E. Peik, and P. O. Schmidt, Rev. Mod. Phys. {\bf87}, 637 (2015).
\bibitem{Cronin}Alexander D. Cronin, J\"{o}rg Schmiedmayer, and David E. Pritchard, Rev. Mod. Phys. {\bf81}, 1051 (2009).
\bibitem{Hammerer}Klemens Hammerer, Anders S. S{\o}rensen, and Eugene S. Polzik, Rev. Mod. Phys. {\bf82}, 1041 (2010).
\bibitem{hu}Jiazhong Hu,∗ Alban Urvoy,∗ Zachary Vendeiro, Valentin Cr\'{e}pel, Wenlan Chen, and Vladan Vuleti\'{c}, Science {\bf358}, 1078 (2017).
\bibitem{Diedrich}F. Diedrich, J. C. Bergquist, W. M. Itano, and D. J. Wineland, Phys. Rev. Lett. {\bf62}, 403 (1989).
\bibitem{Parsons} Maxwell F. Parsons, Florian Huber, Anton Mazurenko, Christie S. Chiu, Widagdo Setiawan, Katherine Wooley-Brown, Sebastian Blatt, and Markus Greiner, Phys. Rev. Lett. {\bf114}, 213002 (2015).
\bibitem{Gröbner} Michael Gr\"{o}bner, Philipp Weinmann, Emil Kirilov, and Hanns-Christoph N\"{a}gerl, Phys. Rev. A {\bf95}, 033412 (2017).
\bibitem{thompson}J. D. Thompson, T. G. Tiecke, A. S. Zibrov, V. Vuleti\'{c}, and M. D. Lukin, Phys. Rev. Lett. {\bf110}, 133001 (2013).
\bibitem{kaufman}A. M. Kaufman, B. J. Lester, and C. A. Regal, Phys. Rev. X {\bf2}, 041014 (2012).
\bibitem{kerman2}Andrew J. Kerman, Ph.D. thesis, Stanford University, 2002.
\bibitem{vuletic}Vladan Vuleti\'{c}, Cheng Chin, Andrew J. Kerman, and Steven Chu, Phys. Rev. Lett. {\bf81}, 5768 (1998).
\bibitem{perrin}H. Perrin, A. Kuhn, I. Bouchoule, and C. Salomon, Europhys. Lett. {\bf42}, 395 (1998).
\bibitem{hamann} S. E. Hamann, D. L. Haycock, G. Klose, P. H. Pax, I. H. Deutsch, and P. S. Jessen, Phys. Rev. Lett. {\bf80}, 4149 (1998).
\bibitem{kerman}Andrew J. Kerman, Vladan Vuleti\'{c}, Cheng Chin, and Steven Chu, Phys. Rev. Lett. {\bf84}, 439 (2000).
\bibitem{han}Dian-Jiun Han, Steffen Wolf, Steven Oliver, Colin McCormick, Marshall T. DePue, and David S. Weiss, Phys. Rev. Lett. {\bf85}, 724 (2000).
\bibitem{Kokkelmans}S. J. J. M. F. Kokkelmans, B. J. Verhaar, K. Gibble, and D. J. Heinzen, Phys. Rev. A {\bf56}, R4389(R) (1997).
\bibitem{li2}Z. Li, S. Singh, T. V. Tscherbul, and K. W. Madison, Phys. Rev. A {\bf78}, 022710 (2008).
\bibitem{papp}S. B. Papp, J. M. Pino, and C. E. Wieman, Phys. Rev. Lett. {\bf101}, 040402 (2008).
\bibitem{altin}P. A. Altin, N. P. Robins, D. Doring, J. E. Debs, R. Poldy, C. Figl, and J. D. Close, Rev. Sci. Instrum. {\bf81}, 063103 (2010).
\bibitem{deh}B. Deh, W. Gunton, B. G. Klappauf, Z. Li, M. Semczuk, J. Van Dongen, and K. W. Madison, Phys. Rev. A {\bf82}, 020701(R) (2010).
\bibitem{cho}Hung-Wen Cho, Daniel J. McCarron, Michael P. K\"{o}ppinger, Daniel L. Jenkin, Kirsteen L. Butler, Paul S. Julienne, Caroline L. Blackley, C. R. Le Sueur, Jeremy M. Hutson, and Simon L. Cornish, Phys. Rev. A {\bf87}, 010703 (2013).
\bibitem{lan}S.-Y. Lan, S. D. Jenkins, T. Chaneli\`{e}re, D. N. Matsukevich, C. J. Campbell, R. Zhao, T. A. B. Kennedy, and A. Kuzmich, Phys. Rev. Lett. {\bf98}, 123602 (2007), and references therein.
\bibitem{lan2} S.-Y. Lan, A. G. Radnaev, O. A. Collins, D. N. Matsukevich, T. A. Kennedy, and A. Kuzmich, Opt. Express {\bf17}, 13639 (2009), and references therein.
\bibitem{zhou}Lin Zhou, Shitong Long, Biao Tang, Xi Chen, Fen Gao, Wencui Peng, Weitao Duan, Jiaqi Zhong, Zongyuan Xiong, Jin Wang, Yuanzhong Zhang, and Mingsheng Zhan, Phys. Rev. Lett. {\bf115}, 013004 (2015), and references therein.
\bibitem{Tai}A. V. Taichenachev, A. M. Tumaikin, V. I. Yudin, and L. Hollberg,
Phys. Rev. A {\bf63}, 033402 (2001).

\bibitem{Belmechri}N. Belmechri, L. F\"{ö}rster, W. Alt, A. Widera, D. Meschede, and A. Alberti, J. Phys. B: At., Mol. Opt. Phys. {\bf46}, 104006 (2013).

\bibitem{Kastberg}A. Kastberg, W. D. Phillips, S. L. Rolston, R. J. C. Spreeuw,
and P. S. Jessen, Phys. Rev. Lett. {\bf74}, 1542 (1995).

\end{thebibliography}

\end{document}